\documentclass[aps,pre,amsmath,amssymb,eqsecnum,amsfonts, reprint,%
amssymb,showkeys,showpacs,bm]{revtex4-1}

\usepackage{graphicx}

\begin{document}

\title{Anomalous scaling in magnetohydrodynamic turbulence: \\
Effects of anisotropy and compressibility in the kinematic approximation}

\author{N.\,V.~Antonov and M.\,M.~Kostenko}

\email{n.antonov@spbu.ru, kontramot@mail.ru}

\affiliation{Chair of High Energy Physics and Elementary Particles \\
Department of Theoretical Physics, Faculty of Physics \\
Saint Petersburg State University, Ulyanovskaja~1 \\
Saint~Petersburg--Petrodvorez, 198904 Russia}

\begin{abstract}
The field theoretic renormalization group and the operator product expansion
are applied to the model of passive vector (magnetic) field advected by a
random turbulent velocity field. The latter is governed by the Navier--Stokes
equation for compressible fluid, subject to external random force with the
covariance $\propto \delta(t-t') k^{4-d-y}$, where $d$ is the dimension
of space and $y$ is an arbitrary exponent. From physics viewpoints, the
model describes magnetohydrodynamic turbulence in the so-called kinematic
approximation, where the effects of the magnetic field on the dynamics of
the fluid are neglected. The original stochastic problem
is reformulated as a multiplicatively renormalizable field theoretic model;
the corresponding renormalization group equations possess an infrared
attractive fixed point. It is shown that various correlation functions
of the magnetic field and its powers demonstrate anomalous
scaling behavior in the inertial-convective range already for small values
of~$y$. The corresponding anomalous exponents, identified with scaling
(critical) dimensions of certain composite fields (``operators'' in the
quantum-field terminology), can be systematically calculated as series
in $y$. The practical calculation is performed in the leading one-loop
approximation, including exponents in anisotropic contributions.
It should be emphasized that, in contrast to Gaussian ensembles with
finite correlation time, the model and the perturbation theory presented
here are manifestly Galilean covariant.
\end{abstract}



\maketitle

\section{Introduction} \label{sec:Intro}

Much attention has been attracted to the problem of intermittency and
anomalous scaling in developed magnetohydrodynamic (MHD) turbulence;
see, e.g.,~\cite{Legacy}--\cite{AG13} and references therein.
It has long been known that in the so-called Alfv{\'e}nic regime, the
MHD turbulence demonstrates the behavior, similar to that of the usual
fully developed fluid turbulence: cascade of energy from the infrared (IR)
range towards smaller scales, where the dissipation effects dominate,
and self-similar (scaling) behavior of the energy spectra in the
intermediate (inertial-convective) range. Moreover, strongly non-Gaussian
(intermittent) character of the fluctuations in the MHD turbulence is much
strongly pronounced than
in ordinary turbulent fluids or in the passive scalar problem.

The solar wind, a conducting fluid expanding into the interplanetary space,
covers a wide range of spatial and temporal scales and thus provides a kind
of ``laboratory'' in which
various models of the MHD turbulence can be tested \cite{Sat}--\cite{Eyink}.
In solar flares, highly energetic and anisotropic large-scale events (with
the magnetic fields as intense as 500 Gauss) coexist with small-scale
stochastic fluctuations and coherent structures, finally responsible for
the dissipation. Thus, modelling the way how the energy is distributed,
conveyed along the scales and eventually dissipated is a difficult task.

The intermittency strongly modifies the IR behavior of the higher-order
correlation functions, leading to anomalous scaling with infinite
sets of independent anomalous exponents \cite{SW2}.

A simplified description of the situation was proposed in~\cite{E}: the
large-scale field $B^{0}_{i} = n_{i} B^{0}$ dominates the dynamics in the
distinguished direction specified by a unit constant vector
${\bf n}=\{n_i\}$, while the fluctuations in the perpendicular
plane are described as nearly two-dimensional. This picture allows for
precise numerical simulations, which show that turbulent fluctuations
organize in rare coherent structures separated by narrow current sheets.
On the other hand, the satellite observations \cite{Sat} and numerical
simulations \cite{GM,SW2} suggest that the scaling behavior in the solar
wind is closer to the anomalous scaling in the three-dimensional fully
developed hydrodynamic turbulence, rather than to simple
Iroshnikov--Kraichnan scaling \cite{Irosh,IK} suggested by the
two-dimensional picture with the inverse energy cascade.

Thus, further analysis of more realistic three-dimensional models is welcome.

In a number of papers, the problem was studied within the framework of the
kinematic approximation, in which the magnetic field is passive in the
sense that it does not affect the dynamics of the velocity field
\cite{V96}--\cite{AG13}. This approximation seems reasonable if the
gradients
of the magnetic fields are not too large. What is more, the renormalization
group analysis of~\cite{K} suggests that such a ``kinematic regime''
can indeed describe the possible IR behavior of the full-scale problem.
It is then possible to model the velocity field ``by hands,'' that is, by
simple statistical ensembles with prescribed properties. Most popular is
the Kazantsev--Kraichnan ensemble \cite{KA68,Kraich2}: the random velocity
field is Gaussian, white in time, and has a power-law spectrum.

Numerous analytical and numerical results were derived for the scalar and
vector fields, advected by the Kazantsev--Kraichnan ``flow,'' see \cite{FGV}
for the review and references. The main results  concerning anomalous
scaling for the magnetic case can be summarized as follows
\cite{V96}--\cite{ABP}:

(i) Anomalous scaling is present and appears already for the
pair correlation function.

(ii) In the presence of large-scale anisotropy (brought about, e.g., by
the constant background field $B^{0}$), the anomalous exponents for a
given correlation function demonstrate a kind of hierarchy: in the expansion
of correlation functions in the spherical harmonics $Y_{lm}$, the
corresponding exponents increase with $l$, the degree of anisotropy.
Thus, for the even-order functions, the leading terms of the inertial-range
behavior are given by the isotropic contribution ($l=0$). This gives
quantitative support for Kolmogorov's hypothesis of the local isotropy
restoration.

(iii) Nevertheless, the anisotropy survives at small scales and manifests
itself in odd-order correlation functions, or in dimensionless ratios
involving such functions (like the skewness factor).

An important advantage of the Kazantsev--Kraichnan ensemble is the
possibility to easily model anisotropy and compressibility.
Importance of compressibility for the MHD turbulence was realized already
in the classical study of \cite{Irosh}. Within the framework of the
Kazantsev--Kraichnan ensemble, effects of compressibility were studied
in \cite{RK97,alpha,J13}. It was shown that:

(iv) The anomalous exponents depend on the degree of compressibility.
When it grows, the hierarchy of anisotropic contributions becomes less
pronounced and the persistence of anisotropy in the depth of the inertial
interval becomes more remarkable.

Of course, generalization of this analysis to more realistic velocity
dynamics is necessary: some of the aforementioned results can be
artefacts of the oversimplified statistics.

It is possible to directly generalize
the Kazantsev--Kraichnan ensemble to the case
of finite correlation time; see, e.g., \cite{A99,AKens,AK2} for the passive
scalar and \cite{TWP} for the passive vector fields. However, such
``synthetic'' models with non-vanishing correlation time suffer from
the lack of Galilean symmetry, which may lead to ``interesting pathologies,''
quoting the authors of~\cite{OU}.
One of such a pathology manifests itself as ultraviolet (UV)
divergence in the vertex \cite{AKens}, which in more realistic models is
forbidden by Galilean invariance.

Thus, it is desirable to describe the advecting velocity field by the
Navier--Stokes equations with a random stirring force and to work
within Galilean covariant formalism. For the incompressible case, the
analysis of the passive vector field was accomplished in~\cite{AG13}.

In this paper, we study the anomalous scaling in the kinematic MHD problem
and model the velocity dynamics by the non-Gaussian velocity field with
finite correlation time, governed by the stochastic Navier--Stokes (NS)
equation. We apply to the problem the approach based on the field theoretic
renormalization group (RG) and the operator product expansion (OPE), earlier
applied to the passive scalar problem \cite{RG,cube}. In that approach, the
anomalous exponents are identified with the critical dimensions of certain
Galilean-invariant composite fields (``operators''). It can be directly
generalized to the cases of finite correlation time, presence of anisotropy,
non-Gaussianity and so on. Passive advection of vector fields (and hence
kinematic MHD problems) with various velocity ensembles were studied earlier
within the RG+OPE approach in~\cite{Lanotte2}--\cite{AG13}.

A general overview of the RG+OPE approach to the problem of anomalous scaling
and more references can be found in \cite{JphysA}.
Detailed exposition of earlier work on the field theoretic RG approach to
stochastic models of turbulence on the whole is presented in \cite{turbo}.

However, analysis of the compressible fluid on the base of stochastic NS
equation appears a difficult task in itself; see,
e.g.,~\cite{Tur}--\cite{ANU}.
In spite of some discrepancies, all of those studies support the
existence of a  ``strongly compressible'' scaling regime,
different from the usual incompressible one.

In the present paper, we adopt the approach of~\cite{ANU}, where, with
the price of some natural approximations, the stochastic NS equation for
a compressible fluid was reformulated as a multiplicatively renormalizable
field theoretic model. Then the standard field theoretic RG was applied to
the problem, and the resulting stationary scaling regime was associated with
the IR attractive fixed point of the corresponding RG equations.

Recently, that ensemble was employed to study, within the RG+OPE framework,
the problem of passive scalar advection in a turbulent compressible fluid
\cite{Kont}. In spite of close resemblance with the case of Kraichnan's
model, some of the results appeared somewhat different. The present
paper continues the study of~\cite{Kont} in connection with the MHD
turbulence. For this reason, we will only briefly discuss the points,
common to the scalar and vector problems, refer the reader to the papers
\cite{ANU,Kont} whenever possible, and focus on the points specific of the
vector case.

The plan of the paper is the following.
In section~\ref{sec:Model} we give the detailed description of the model:
the velocity ensemble, the stochastic MHD equation and the field theoretic
formulation. In section~\ref{sec:Reno} we discuss canonical dimensions and
renormalization of the field theoretic model, demonstrate its multiplicative
renormalizability and calculate (in the leading one-loop approximation)
the corresponding renormalization constant. In section~\ref{sec:RGE} we
derive the corresponding RG equations and show that they possess the only
IR attractive fixed point in the physical region of parameters. This fact
implies the scaling behavior in the IR range (long times, large distances);
the corresponding critical dimensions of the basic fields and parameters
are presented. In section~\ref{sec:Opera} we calculate the critical
dimensions of the tensor composite fields (operators), constructed
solely of the basic scalar
fields; these will play the crucial role in the following.
In section~\ref{sec:OPE} we employ the OPE to derive the inertial-range
asymptotic behavior of various correlation functions.
Section~\ref{sec:Conc} is reserved for discussion and the conclusion.

\section{The Model} \label{sec:Model}

\subsection{The Velocity Ensemble} \label{sec:ANU}

Following \cite{ANU,Kont}, we describe the stochastic dynamics of a
compressible fluid by the set of two equations:
\begin{eqnarray}
\nabla_{t} v_{i} &=&
\nu_{0} [\delta_{ik}\partial^{2}-\partial_{i}\partial_{k}]
v_{k}\! +\! \mu_0 \partial_{i}\partial_{k} v_{k} -\!
\partial_{i} \phi\! +\! f_{i}
\label{ANU} \\
\nabla_{t} \phi &=& -c_{0}^{2} \partial_{i}v_{i},
\label{ANU1}
\end{eqnarray}
which are derived from the momentum balance equation and the continuity
equation \cite{LL}
with two assumptions: the kinematic viscosity coefficients
$\nu_{0}$ and $\mu_{0}$ are assumed to be constant, that is, independent of
$x=\{t, {\bf x}\}$,
and the equation of state is taken in the simplest form of the linear
relation $(p-\bar p) = c^{2}_{0} (\rho-\bar\rho)$
between the deviations of the pressure $p(x)$ and the
density $\rho(x)$ from their mean values; then the constant $c_{0}$
has the meaning of the (adiabatic) speed of sound.

In Eqs. (\ref{ANU}), (\ref{ANU1}),
$\boldsymbol{v} = \{v_i (x)\}$ is the velocity field and,
instead of the density, we use the scalar field defined as
$\phi(x) = c_{0}^{2} \ln (\rho(x)/\bar \rho)$. Furthermore,
\begin{eqnarray}
\nabla_{t} = \partial_{t} + v_{k} \partial_{k}
\label{Nabla}
\end{eqnarray}
is the Lagrangean (Galilean covariant) derivative, $\partial_{t} =
\partial /\partial t$, $\partial_{i} = \partial /\partial x_{i}$, and
$\partial^{2} =\partial_{i}\partial_{i}$ is the Laplace operator.
The problem is studied in the
$d$-dimensional (for generality) space ${\bf x}=\{x_i\}$, $i=1\dots d$,
and the summations over the repeated Latin indices are always implied.

In the Navier--Stokes equation (\ref{ANU}), $f_{i}$ is the density of the
external force (per unit mass), which mimics the energy input into the
system from the large-scale stirring. In order to apply the standard
perturbative RG to the problem, it is taken to be Gaussian with zero mean,
not correlated in time (this is dictated by the Galilean symmetry),
with the given covariance
\begin{eqnarray}
\langle f_{i}(x) f_{j}(x') \rangle = \delta(t-t') \int_{k>m} \frac{d{\bf k}}
{(2\pi)^{d}} \, D^{f}_{ij}({\bf k}) \exp\{{\rm i} {\bf k}\cdot{\bf x}\},
\nonumber \\ {}
\label{force}
\end{eqnarray}
with
\begin{eqnarray}
D^{f}_{ij}({\bf k}) = D_{0}\, k^{4-d-y}\,
\left\{ P^{\bot}_{ij} ({\bf k})
+ \alpha P^{\parallel}_{ij} ({\bf k}) \right\}.
\label{power}
\end{eqnarray}
Here $P^{\bot}_{ij} ({\bf k})=\delta_{ij}-k_{i}k_{j}/k^{2}$ and
$P^{\parallel}_{ij} ({\bf k})=k_{i}k_{j}/k^{2}$ are the transverse and the
longitudinal projectors, respectively, $k=|{\bf k}|$ is the wave number
(momentum), $D_{0}$ and $\alpha$ are positive amplitudes.
The parameter $g_{0}=D_{0}/\nu_0^{3}$ plays the part of the coupling
constant (expansion parameter in the perturbation theory);
the relation $g_{0} \sim \Lambda^{y}$ defines the typical UV momentum scale.
The parameter $m \sim L^{-1}$, reciprocal of the integral turbulence scale,
provides IR regularization; its precise form is unessential and the sharp
cut-off is merely the simplest choice for calculational reasons.
The exponent $0<y\le 4$ plays the part analogous to $\varepsilon=4-d$
in the RG theory of critical state \cite{Zinn,Book3}: it provides UV
regularization (so that the UV divergences have the form of the poles in
$y$) and various scaling dimensions are calculated as series in $y$.
The most realistic (physical) value is given by the limit $y\to4$: then
the function (\ref{power}) can be viewed as a power-like representation
of the function $\delta({\bf k})$ and it corresponds to the idealized
picture of the energy input from infinitely large scales.

As already mentioned, more detailed justification and
discussion of the compressible model (\ref{ANU})--(\ref{power})
is given in \cite{ANU,Kont}.

\subsection{The MHD Equation} \label{sec:MHD}

In the presence of a constant background field $B^{0}_{i}=B^{0}n_{i}$ with
a certain constant unit vector ${\bf n}=\{n_i\}$, the
dynamic equation for the fluctuating part $\theta_{i}=\theta_{i}(t,{\bf x})$
of the full magnetic field $B_{i} = B^{0}(n_{i}+ \theta_{i})$ has the form
\begin{eqnarray}
\partial _t\theta_{i} + \partial_{k} (v_{k}\theta _{i} - \theta_{k}
v_{i}) = \kappa_{0}\partial^{2} \theta_{i} + n_{k}\partial_{k} v_{i},
\label{mhd}
\end{eqnarray}
where $\kappa_{0}= c_{l}^{2}/4\pi\sigma$ is the magnetic diffusion
coefficient.
Equation (\ref{mhd}) follows from the Maxwell equations neglecting the
displacement current and the simplest form of Ohm's law for a moving medium
${\bf j}=\sigma \left({\bf E} + c_{l}^{-1}[\boldsymbol{v},{\bf B}]\right)$,
where $\sigma$ is the conductivity and $c_{l}$ is the speed of light;
see, e.g., \cite{Moffat}.

The last term on the right-hand side of (\ref{mhd}) maintains the steady
state of the system and acts as the source of the anisotropy; in principle,
it can be replaced with an artificial Gaussian noise with appropriate
statistics. In the real problem, the field $\boldsymbol{v}$ satisfies the
Navier--Stokes equation with the additional Lorentz force term
$\sim ({\bf B} \times {\rm curl}\,{\bf B})$. In our kinematic approximation
the back reaction of the magnetic field on the velocity dynamics is neglected
and the latter is described by the stochastic problem
(\ref{ANU})--(\ref{power}) without the Lorentz term.

\subsection{The field theoretic formulation} \label{sec:QFT}

It is well known that any stochastic problem of the type
(\ref{ANU})--(\ref{power}) can be reformulated, in a standard fashion,
as a certain field theoretic
model; see, e.g., \cite{Zinn,Book3}. This means that various correlation
and response
functions of the original stochastic problem can be represented as functional
integrals over the doubled set of fields $\Phi$ with the weight
$\exp {\cal S}(\Phi)$, where ${\cal S}(\Phi)$ is the so-called
De Dominicis--Janssen action functional. The action functional
${\cal S}_{v}(\Phi)$ for the problem (\ref{ANU})--(\ref{power}) with
$\Phi=\{\boldsymbol{v}',  \phi', \boldsymbol{v}, \phi\}$ looks too
cumbersome, and we do not reproduce it here, as well as the elements of
the corresponding Feynman diagrammatic techniques (bare propagators and
vertices); they can be found in~\cite{ANU,Kont}. Below we only will
need the velocity--velocity propagator at $c_{0}=0$; in the
frequency--momentum ($\omega$--${\bf k}$) representation it has the form:
\begin{equation}
\langle v_{i}v_{j}\rangle_{0}   = D_{0} \left\{
\frac {P^{\bot}_{ij} ({\bf k})}{ \omega^{2} + \nu^2_{0} k^{4}}
+ \frac {\alpha P^{\parallel}_{ij} ({\bf k})}
{ \omega^{2} + u^{2}_{0} \nu^2_{0} k^{4}} \right\} .
\label{lines}
\end{equation}

The full-scale stochastic problem (\ref{ANU})--(\ref{mhd}) corresponds
to the action functional
\begin{equation}
{\cal S}(\Phi)={\cal S}_{v} (\boldsymbol{v}',  \phi', \boldsymbol{v}, \phi)
+ {\cal S}_{\theta}( \boldsymbol{\theta}', \boldsymbol{\theta},
\boldsymbol{v}),
\label{FSA}
\end{equation}
where
\begin{equation}
{\cal S}_{\theta} = \theta'_{i} \bigl\{
-\partial _t\theta_{i} - \partial_{k} (v_{k}\theta _{i} - \theta_{k} v_{i})
+  \kappa_{0}\partial^{2} \theta_{i} + n_{k}\partial_{k} v_{i}\bigr\}
\label{MHDA}
\end{equation}
is the De Dominicis--Janssen action for the problem (\ref{mhd}) at fixed
$\boldsymbol{v}$. It brings about the new propagator
\begin{equation}
\langle \theta'_{i}\theta_{j} \rangle_{0}   =
\frac{P^{\bot}_{ij} ({\bf k})}
{-{\rm i}\omega+ w_{0} \nu_{0} k^{2} }
\label{lines3}
\end{equation}
and the new vertex $V_{ijl}\theta'_{i}\theta_{j}v_{l}$
with the vertex factor
\begin{equation}
V_{ijl}({\bf k}) = {\rm i} (\delta_{ij}k_{l}-\delta_{il}k_{j}).
\label{vertex1}
\end{equation}
A few remarks are in order here. First, the derivative at the vertex in
(\ref{MHDA}) can be moved onto the auxiliary field~$\boldsymbol{\theta}'$
using the integration by parts; thus ${\bf k}$ in (\ref{vertex1}) is the
momentum argument of $\boldsymbol{\theta}'$. Second, the vertex factor
satisfies the transversality condition
\begin{equation}
k_{i}V_{ijl}({\bf k})=0,
\label{trans1}
\end{equation}
that follows from its explicit form (\ref{vertex1}). It remains to note that
another new mixed propagator $\langle \theta\boldsymbol{v}\rangle_{0}$ will
not appear in relevant diagrams.

\section{UV divergences and the Renormalization}
\label{sec:Reno}

\subsection{Canonical dimensions, counterterms, and renormalizability}
\label{sec:Canon}

The analysis of UV divergences is based on the
analysis of canonical dimensions; see, e.g., \cite{Zinn,Book3}.
Dynamical models of the type (\ref{ANU})--(\ref{mhd})
have two scales: the time scale $T$ and the length scale $L$.
The canonical dimension of any quantity $F$ (a field or a parameter) is
described by two numbers, the momentum dimension $d_{F}^{k}$
 and the
frequency dimension $d_{F}^{\omega}$,
defined such that $[F] \sim [T]^{-d_{F}^{\omega}} [L]^{-d_{F}^{k}}$.
In the renormalization theory, the central role is played by the
total canonical dimension  $d_{F}=d_{F}^{k}+2d_{F}^{\omega}$, defined such
that all the viscosity or diffusivity coefficients are dimensionless;
see \cite{turbo,Book3}. All the canonical dimensions in our model
(\ref{ANU})--(\ref{mhd}) are identical to their counterparts in the scalar
case, and we refer the reader to the Table~1 in~\cite{Kont}.

The formal index of UV divergence of a certain 1-irreducible Green function
$\Gamma$ is given by its total canonical dimension :
\begin{eqnarray}
\delta_{\Gamma} = d+2 - \sum_{\Phi} N_{\Phi} d_{\Phi},
\label{index}
\end{eqnarray}
where $N_{\Phi}$ are the numbers of the fields entering into the function
$\Gamma$, $d_{\Phi}$ are their total canonical dimensions.
Superficial UV
divergences, whose removal requires counterterms, can be present only in
the functions $\Gamma$ with a non-negative integer $\delta_{\Gamma}$.
The counterterm is a polynomial in frequencies and momenta of degree
$\delta_{\Gamma}$, with the convention that $\omega \sim k^{2}$.

The dimensional analysis (``power counting'') should be augmented by
certain additional considerations:

(i) All the 1-irreducible Green functions without the auxiliary (``primed'')
fields vanish identically and thus require no counterterms.

(ii) If a number of external momenta occurs as an overall
factor in all the diagrams of a certain Green function, the real index of
divergence $\delta_{\Gamma}'$ is smaller than $\delta_{\Gamma}$ by the
corresponding number of unities. In the model ${\cal S}_{v}$ the field
$\phi$ enters the corresponding vertex only in the form of a spatial
derivative, which reduces the real index of divergence:
\begin{equation}
\delta_{\Gamma}' = \delta_{\Gamma}- N_{\phi}.
\label{real}
\end{equation}

(iii) The Galilean invariance of the model requires that the counterterms
be also invariant. In particular,
this means that the covariant derivative (\ref{Nabla}) appears in the
counterterms as a whole.

These considerations forbid superficial UV divergences in certain Green
functions, allowed by dimensional analysis, and hence reduce the number
of the counterterms needed for the renormalization of the model.

The analysis of the field theoretic model with the action ${\cal S}_{v}$ in
(\ref{FSA}), performed in \cite{ANU} (see also \cite{Kont}), has shown that
it is multiplicatively renormalizable (after a simple natural extension).
This means that all the UV divergences can be removed from the Green
functions by the renormalization of the fields $\phi\to Z_{\phi}\phi$,
$\phi'\to Z_{\phi'}\phi'$ and of the parameters:
\begin{equation}
g_{0} = g\mu^y Z_{g}, \quad \nu_0 Z_{\nu}, \quad c_{0} = c Z_{c}\, ,
\label{Ren}
\end{equation}
and so on. Here the renormalization constants $Z_{i}$ absorb all the UV
divergences, so that the Green functions are UV finite (that is, finite at
$y=0$) when expressed in terms of the renormalized parameters $g,u$, and so
on; the reference scale (or the ``renormalization mass'') $\mu$ is an
additional free parameter of the renormalized theory. No renormalization
of the fields $\boldsymbol{v}',\boldsymbol{v}$
and of the parameters $m,\alpha$ is needed.

The inclusion of the new contribution ${\cal S}_{\theta}$ in the full model
brings about the only new UV divergence in the 1-irreducible function
$\langle \theta'\theta \rangle_{1-ir}$ with the counterterm
$\theta'\partial^{2}\theta$. Two points are important here:

(iv) From the linerarity  of the original stochatic model in the field
$\theta$ it follows that $N_{\theta'}- N_{\theta}$ is a non-negative integer
for any nontrivial 1-irreducible Green function: no other Feynman diagram
can be drawn. This fact forbids the superficial divergences in all
the 1-irreducible functions
$\langle \theta'\theta\dots \theta\rangle_{1-ir}$, except for the first one,
and thus prevents our model from being non-renormalizable, despite the fact
that the magnetic field has a negative canonical dimension.

(v) For the full model (\ref{FSA}), the items (ii) and (iii) require
some additional discussion.
The derivative at the vertex in ${\cal S}_{\theta}$ can be moved,
using the integration by parts, onto the field $\theta'$. Thus, the real
index of divergence is reduced according to the item (ii) above, and
$\theta'$ enters the countertems only as a spatial derivative.
The expression (\ref{real}) has to be replaced with
\begin{equation}
\delta_{\Gamma}' = \delta_{\Gamma}- N_{\phi}- N_{\theta'}.
\label{real2}
\end{equation}
Thus, the counterterm $\theta'\partial_{t}\theta$ is forbidden, and so is
$\theta'(v_{i}\partial_{i})\theta$ due to the Galilean symmetry (iii).

The only remaining counterterm $\theta'\partial^{2}\theta$ is naturally
reproduced by the multiplicative renormalization of the magnetic diffusion
coefficient: $\kappa_{0}=\kappa Z_{\kappa}$. No renormalization of the
fields ${\theta}'$, ${\theta}$ is needed.

The renormalized analog of the action functional (\ref{FSA}) has the form
\begin{equation}
{\cal S}^{R}(\Phi)={\cal S}_{v}^{R}(\Phi) +
{\cal S}_{\theta}^{R}(\Phi),
\label{FactR}
\end{equation}
where
${\cal S}^{R}(\Phi)$ is the renormalized analog of the action
${\cal S}(\Phi)$, given in \cite{ANU,Kont}, and
\begin{equation}
{\cal S}_{\theta}^{R} = \theta'_{i} \bigl\{
-\partial _t\theta_{i} - \partial_{k} (v_{k}\theta _{i} - \theta_{k} v_{i})
+ \kappa Z_{\kappa} \partial^{2} \theta_{i} +
n_{k}\partial_{k} v_{i}\bigr\}
\label{Rak}
\end{equation}
is the renormalized part of the full action that describes interaction
with the magnetic field.

\subsection{Leading-order calculation of the renormalization constant
$Z_{\kappa}$} \label{sec:Sigma}

We performed the practical calculation of the renormalization constant
$Z_{\kappa}$ in the leading one-loop approximation, which is consistent with
the accuracy of the calculation for the NS problem (\ref{ANU}), (\ref{ANU1})
made in \cite{ANU}. Although this calculation is rather simple and similar to
that for the Gaussian velocity ensemble \cite{TWP}, we will briefly discuss
it for the sake of completeness and in order to stress some peculiarities.

The constant is found from the requirement that the 1-irreducible Green
function $\langle\theta'\theta\rangle_{\rm 1-ir}$ be UV finite (that is,
finite at $y\to0$) when expressed in renormalized parameters. In the
frequency--momentum representation it has the form:
\begin{equation}
\langle\theta'_{1}\theta_{2}\rangle_{\rm 1-ir}(\Omega,{\bf p}) =
\left\{- \kappa_0 p^{2} + {\rm i}\Omega  \right\}
\, P_{12}^{\bot}({\bf p})
+ \Sigma_{12} (\Omega,  {\bf p}),
\label{Dyson}
\end{equation}
where  $\Sigma_{12}$ is the ``self-energy operator'' given by infinite sum
of 1-irreducible Feynman diagrams and $p=|{\bf p}|$.
Because of the large number of tensor
indices involved in our expressions, we use numbers (instead of latin
letters) to denote them, with the standard convention on the summation over
repeated indices.

The only one-loop self-energy diagram looks as follows:
\begin{equation}
\Sigma_{12} = 
\includegraphics[width=.13\textwidth,clip]{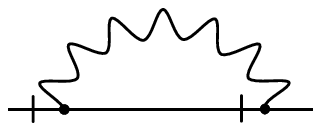} .
\label{Self}
\end{equation}
Here the wavy line denotes the bare propagator $\langle vv\rangle_0$, the
solid line with a slash
denotes the bare propagator $\langle \theta \theta' \rangle _0$ from
(\ref{lines3}), the slashed end corresponding to the field $\theta'$.
The dots with three attached fields $\theta'$, $\theta$, $v$ denote the
vertex (\ref{vertex1}).

In this approximation, the renormalization constant in the bare term of
(\ref{Dyson}) is taken to the first order in $g$, while in the diagram
(\ref{Self}) all $Z$'s are simply replaced with unities.
Furthermore, we only need to know the divergent part of (\ref{Self}), which
is quadratic in ${\bf p}$ (see the preceding subsection). Thus, we can put
$\Omega=0$ in (\ref{Dyson}) and retain only quadratic terms
in the expansion of $\Sigma_{12} (\Omega=0,  {\bf p})$
in ${\bf p}$. Like for the original NS model,
its divergent part is independent of $c_{0} \sim c$ and can be calculated
directly at $c=0$; see the discussion in \cite{Kont}. Thus, we can use the
expression (\ref{lines}) for $\langle vv\rangle_0$.

Then the analytic expression for (\ref{Self}) takes on the form:
\begin{widetext}
\begin{equation}
\Sigma_{12} (\Omega=0, {\bf p})  = D_{0}
\int\frac{d\omega}{2\pi}
\int_{k>m}\frac{d{\bf k}}{(2\pi)^{d}}\,
V_{143}({\bf p}) V_{625}({\bf p}+{\bf k})
\left\{
\frac {P^{\bot}_{35} ({\bf k})}{ \omega^{2} + \nu^2 k^{4}}
+ \frac {\alpha P^{\parallel}_{35} ({\bf k})}
{ \omega^{2} + u^{2} \nu^2 k^{4}} \right\}\,
\frac{P^{\bot}_{46}({\bf k}+{\bf p}) }
{-{\rm i}\omega +w\nu |{\bf k}+{\bf p}|^{2}},
\label{pesez}
\end{equation}
\end{widetext}
where $k=|{\bf k}|$.
The simplifying replacement $P^{\bot}_{46}\to\delta_{46}$ can immediately
be made owing to the transversality condition (\ref{trans1}) and the
contraction with $V_{625}$.

Integrations over the frequency are easily performed, for example:
\begin{widetext}
\begin{equation}
\int\frac{d\omega}{2\pi}\,
\frac{1}{-{\rm i}\omega+w\nu |{\bf p}+{\bf k}|^{2}}\,
\frac {1} { \omega^{2} + u^{2}\nu^2 k^{4}} =
 \frac {1} {2u\nu^{2}k^{2}(uk^{2}+w|{\bf p}+{\bf k}|^{2})}.
\label{Fre}
\end{equation}
\end{widetext}

The numerators in the integrand of (\ref{pesez}) contain the terms
quadratic and linear in ${\bf p}$. For the first ones,
one can immediately set ${\bf p}=0$ in (\ref{Fre}), while for the
second ones, one should expand (\ref{Fre}) up to the linear term in
${\bf p}$, for example,
\[ \frac {1} {uk^{2}+w|{\bf p}+{\bf k}|^{2}} =
\frac{1}{(u+w)k^{2}} \left\{ 1- \frac{2w}{(u+w)} \frac{(\bf pk)}{k^{2}}
\right\}. \]
With the aid of the formulas
\begin{eqnarray}
\int\! {d{\bf k}} k_{i} f(k) =0, \quad
\int\! {d{\bf k}} \frac{k_{i}k_{s}}{k^{2}} f(k) =
\frac{\delta_{is}}{d}\, \int {d{\bf k}}\, f(k), \
\nonumber \\
\int\! {d{\bf k}} \frac{k_{i}k_{s}k_{l}k_{p}}{k^{4}} f(k) =
\frac{\delta_{is}\delta_{lp}+\delta_{il}\delta_{sp}+\delta_{ip}\delta_{sl}}
{d(d+2)} \int {d{\bf k}} f(k),
\nonumber \\ {}
\label{tenz}
\end{eqnarray}
where $f(k)$ is any function depending only on $k=|{\bf k}|$, all the
resulting integrals are reduced to the scalar integral
\begin{eqnarray}
J(m)= \int_{k>m} {d{\bf k}}\, \frac{1}{k^{d+y}} = S_{d}
\frac{m^{-y}}{y}\, ,
\label{ska}
\end{eqnarray}
where
\begin{equation}
S_d= 2\pi^{d/2}/\Gamma (d/2)
\label{Sd}
\end{equation}
is the surface area of the unit sphere in the $d$-dimensional space and
$\Gamma(\cdots)$ is Euler's Gamma function.

The final result contains two types of terms, proportional to
$P^{\bot}_{12} ({\bf p})$ and $P^{\parallel}_{12} ({\bf p})$, respectively.
Owing to the transversality of the fields $\theta$, $\theta'$, the latter
ones should be discarded. (This would happen automatically if we included
the corresponding projector into the vertex (\ref{vertex1}), but we did
not do that for brevity). Practically, it is more convenient to collect
only terms proportional to $\delta_{12}\,p^{2}$ and drop all the terms
proportional to $p_{1}p_{2}$ in the course of calculation.
It remains to express the amplitude $D_{0}$ in (\ref{pesez}) in terms of
renormalized variables: $D_{0}=g\nu^3 \mu^y$.

Then the final result reads:
\begin{widetext}
\begin{equation}
\Sigma_{12} (\Omega=0,  {\bf p}) = - \nu p^{2}\,P^{\bot}_{12} ({\bf p})\,
\frac{\hat g}{2dy}\, \left(\frac{\mu}{m}\right)^{y}\, \left\{
\frac{(d-1)}{1+w} + \frac{\alpha(u-w)}{u(u+w)^{2}} \right\}.
\label{DysonF}
\end{equation}
\end{widetext}
Here we passed to the new coupling constant
\begin{eqnarray}
{\hat g}=g S_{d}/(2\pi)^{d},
\label{ghat}
\end{eqnarray}
with $S_{d}$ from (\ref{Sd}).

Then in the MS scheme the renormalization constant $Z_{\kappa}$ that
cancels the pole of the expression (\ref{DysonF}) in the renormalized
analog of the function  (\ref{Dyson}) (that is, with the replacement
$\kappa_{0}\to \kappa Z_{\kappa}$ in the bare term) has the form:
\begin{equation}
Z_{\kappa} = 1 - \frac{{\hat g}} {2dwy}
\left\{ \frac{(d-1)}{(1+w)} + \frac{\alpha(u-w)}{u(u+w)^{2}}
\right\},
\label{Zk}
\end{equation}
while the corresponding anomalous dimension is
\begin{equation}
\gamma_{\kappa} =  \frac{\hat g} {2dw}
\left\{ \frac{(d-1)}{(1+w)} + \frac{\alpha(u-w)}{u(u+w)^{2}}
\right\},
\label{gk}
\end{equation}
with the corrections of the order ${\hat g}^{2}$ and higher.

It is interesting to note that the expression (\ref{gk}) literally coincides
with its analog for the scalar fields advected by the same velocity ensemble;
see Eq.~(3.24) in~\cite{Kont}.
Similar coincidence between the passive scalar and magnetic fields in the
kinematic approximation was earlier observed for the incompressible case
(see, e.g.,~\cite{turbo});  sometimes it extends to the two-loop
approximation~\cite{Marian}.

\section{RG equations, fixed point and the critical dimensions}
\label{sec:RGE}

Here we only briefly discuss the derivation of the IR scaling behavior from
the RG equations in our model; it is nearly identical to the scalar case,
discussed in \cite{Kont} in great detail.

Multiplicative renormalizability of the field theoretic model (\ref{FSA})
allows one to derive, in a standard way, differential RG equations for the
renormalized Green functions
\[ G(e,\mu,\dots) = \langle \Phi \dots \Phi \rangle_{R}. \]
Here
$e=\{g,\nu,u,v,w,c,m,\alpha\}$ is the full set of renormalized parameters,
$\mu$ is the reference momentum scale and the ellipsis stands for the other
arguments (times or frequencies and coordinates or momenta). For convenience,
we introduced here three dimensionless ratios: $u_{0}=\mu_0/\nu_0$ and
$v_{0}=\chi_0/\nu_0$ are related to the viscosity and diffusivity coefficients
of the (properly extended) model (\ref{ANU}), (\ref{ANU1}), while
$w_{0}=\kappa_0/\nu_0$ is related to the magnetic diffusivity coefficient;
$u,v,w$ are their renormalized analogs.

The RG equation expresses the invariance of the renormalized Green function
with respect to changing of the reference scale $\mu$, when the bare
parameters $e_{0}$ are kept fixed:
\begin{equation}
\left\{ \widetilde{\cal D}_{\mu} + \sum_{\Phi} N_{\Phi}\gamma_{\Phi}
\right\} \,G(e,\mu,\dots) = 0.
\label{RGE}
\end{equation}
Here and below we denote ${\cal D}_{x} \equiv x\partial_{x}$ for any
variable $x$ and $\widetilde{\cal D}_{\mu}$ is the operation
${\cal D}_{\mu} \equiv \mu\partial_{\mu}$ at fixed $e_{0}$. In terms of
the renormalized variables, it takes the form
\begin{equation}
\widetilde{\cal D}_{\mu}  = {\cal D}_{\mu} + \beta_{g}\partial_{g} +
\beta_{u}\partial_{u} + \beta_{v}\partial_{v}+ \beta_{w}\partial_{w}
- \gamma_{\nu}{\cal D}_{\nu}- \gamma_{c}{\cal D}_{c} .
\label{RG2}
\end{equation}
The anomalous dimension $\gamma_{F}$ of a certain quantity $F$
(a field or a parameter) is defined by the relation
\begin{equation}
\gamma_{F}= \widetilde{\cal D}_{\mu} \ln Z_F ,
\label{RGF1}
\end{equation}
and the $\beta$ functions for the dimensionless parameters (``coupling
constants'') are
\begin{eqnarray}
\beta_{g} &=& \widetilde{\cal D}_{\mu} g = g\,[-y-\gamma_{g}],
\nonumber \\
\beta_{u} &=& \widetilde{\cal D}_{\mu} u = -u\gamma_{u},
\label{betagw}
\end{eqnarray}
and similarly for $\beta_{v}$, $\beta_{w}$. Here the second equalities
result from the definitions and the relations of the type (\ref{Ren}).

Note that from the definition of $w_{0}$ it follows that
$Z_{\kappa}=Z_{\nu}Z_{w}$, so that
\begin{eqnarray}
\beta_{w} =  w [\gamma_{\nu} - \gamma_{w}].
\label{betaw}
\end{eqnarray}

The possible types of IR asymptotic behavior are associated with IR
attractive fixed points of the RG equations. The coordinates
$g_{*}=\{g_{i*}\}$ of the fixed points are found from the equations
\begin{equation}
\beta_{i} (g_{*}) =0,
\label{points}
\end{equation}
where $g=\{g_i\}$ is the full set of coupling constants and
$\beta_{i}=\widetilde{\cal D}_{\mu}g_i$ are their $\beta$ functions.
The character of a fixed point is determined by the matrix
\begin{equation}
\Omega_{ij}=\partial\beta_{i}/\partial g_{j}|_{g=g_{*}}.
\label{Omega}
\end{equation}
For the IR fixed points the matrix $\Omega$ is positive
(that is, positive are the real parts of all its eigenvalues).

The analysis performed in~\cite{ANU} (see also \cite{Kont}) on the base
of the leading-order (one-loop) approximation has shown that the RG
equations of the model ${\cal S}_{v}$, corresponding to the stochastic NS
problem (\ref{ANU}), (\ref{ANU1}),
possess the only IR attractive fixed point in the physical region of
parameters ($g,u,v>0$):
\begin{equation}
\hat g_{*} = \frac{4dy}{3(d-1)} +O(y^{2}), \quad u_{*}=1+O(y), \quad
v_{*}=1+O(y).
\label{FP}
\end{equation}

From a certain exact relation between the renormalization constants
\cite{ANU}, the exact result
\begin{eqnarray}
\gamma_{\nu}^{*}=y/3
\label{Anom}
\end{eqnarray}
follows (no corrections of the order $y^{2}$ and higher). Here and below,
$\gamma_{i}^{*}$ denotes the value of the anomalous dimension
$\gamma_{i}$ at the fixed point.

Now we substitute the one-loop expressions (\ref{gk}), (\ref{FP})
and the exact result (\ref{Anom}) into Eq. (\ref{betaw}). Then the
equation $\beta_{w} =0$ yields, after simple algebra, the equation
\begin{equation}
(w-1) [(d-1)(w+1)(w+2)+2\alpha] =0,
\label{kub}
\end{equation}
which has the only positive solution $w_{*}=1$, with possible corrections
of order $O(y)$ and higher.

Since the functions $\beta_{g,u,v}$ do not depend on $w$, the new eigenvalue
of the matrix (\ref{Omega}) coincides with the diagonal element
\[\partial\beta_{w}/\partial w|_{g=g_{*}} = y\,
\{3(d-1)+\alpha\}\, / 6(d-1)>0; \]
thus the fixed point (\ref{FP}) and $w_{*}=1$ of the full model
remains IR attractive.

Existence of an IR attractive fixed point in the physical region of the
parameters implies existence of scaling behavior in the IR range.
The critical dimension of some quantity $F$ (a field or a parameter)
is given by the relation (see \cite{turbo,Book3})
\begin{equation}
\Delta_{F} = d^{k}_{F}+ \Delta_{\omega}d^{\omega}_{F} + \gamma_{F}^{*},
\quad
\Delta_{\omega} =  2-\gamma_{\nu}^{*} = 2-y/3.
\label{Krit}
\end{equation}
Here $d^{k}_{F}$ and $d^{\omega}_{F}$ are the canonical dimensions of $F$,
$\gamma_{F}^{*}$ is the value of the anomalous dimension $\gamma_{F}$
at the fixed point,
and $\Delta_{\omega}$ is the critical dimension of the frequency.

The critical dimensions of the fields and parameters of the model described
by the action ${\cal S}_{v}$ from Eq. (\ref{FSA}) are presented
in~\cite{ANU}; see also ~\cite{Kont}. In addition, our full model involves
two more critical dimensions:
\begin{equation}
\Delta_{\theta}= -1+y/6, \quad \Delta_{\theta'}= d+1 -y/6.
\label{KriTet}
\end{equation}
These expressions are exact because the fields ${\theta}$ and
${\theta}'$ are not renormalized.

\section{Composite fields and their dimensions} \label{sec:Opera}

An important role in the following will be played by certain composite
fields (``composite operators'' in the quantum-field terminology). In
general, a local composite operator is a monomial or polynomial built of
the primary fields $\Phi(x)$ and their finite-order derivatives at a single
space-time point $x$. In the Green functions with such objects, new UV
divergences arise due to coincidence of the field arguments. They should be
eliminated by additional renormalization procedure. As a rule, operators mix
in renormalization: renormalized operators (whose Green functions are UV
finite) are given by finite sums of the original monomials. However, in the
following only a simpler situation will be encountered, when the original
operator $F(x)$ and its renormalized analog $F^{R}(x)$ are related by
multiplicative renormalization $F(x)= Z_{F} F^{R}(x)$ with a single
renormalization constant $Z_{F}$. Then the critical dimension $\Delta_{F}$
of the operator~$F$ is given by the same expression (\ref{Krit}) and, in
general, differs from the naive sum of the dimensions of the fields and
derivatives that compose the operator.

We will focus on the irreducible tensor operators built solely of the fields
$\theta$. They have the forms
\begin{equation}
F_{nl}\equiv \theta_{i_{1}}(x)\cdots \theta_{i_{l}}(x)\,
\left(\theta_{i}(x)\theta_{i}(x)\right)^{s} + \dots,
\label{Fnl}
\end{equation}
where $l\le n$ is the number of free tensor indices and $n=l+2s$ is
the total number of the fields $\theta$ entering into the
operator; the tensor indices and the argument $x$ of the symbol
$F_{nl}$ are omitted. The ellipsis stands for the appropriate
subtractions involving the Kronecker delta symbols, which ensure
that the resulting expressions are traceless with respect to
contraction of any given pair of indices, for example,
$\theta_{i}\theta_{j} - \delta_{ij}(\theta_{k}\theta_{k}/d)$ and so on.

The total canonical dimension of any 1-irreducible Green function $\Gamma$
with one operator $F(x)$ and arbitrary number of primary fields
(the formal index of UV divergence) is  given by
\begin{eqnarray}
\delta_{\Gamma} = d_{F} - \sum_{\Phi} N_{\Phi} d_{\Phi},
\label{indexo}
\end{eqnarray}
where $N_{\Phi}$ are the numbers of the fields entering into $\Gamma$,
$d_{\Phi}$ are their total canonical dimensions, and $d_{F}$ is the
canonical dimension of the operator. Superficial UV divergences can be
present only in the functions
$\Gamma$ with a non-negative integer $\delta_{\Gamma}$. For the operators
(\ref{Fnl}) from Table~1 in~\cite{Kont} we find $d_{F}=-n$. The
linearity of the equation (\ref{mhd}) in the field $\theta$ imposes the
restriction that $N_{\theta}$ in (\ref{indexo}) cannot exceed the number
of the fields $\theta$ in the operator $F$. The direct analysis shows that
superficial UV divergences $(\delta_{\Gamma}\ge0)$ for $F_{nl}$ can be
present only in the 1-irreducible functions with $N_{\theta'}=N_{v}=0$ and
$N_{\theta}=n$; they are all logarithmic: $\delta_{\Gamma} =0$. The simple
inspection shows that the mixed propagator $\langle\theta v\rangle_{0}$
does not appear in the relevant Feynman diagrams; in other words, the last
term in the right-hand side of the equation (\ref{mhd}) is unimportant
here. Without that term, the model becomes $O(d)$ invariant. In turn, this
means that irreducible operators with different values of $l$ cannot mix
with each other. We finally conclude that the operators (\ref{Fnl})
renormalize multiplicatively: $F_{nl} = Z_{nl} F_{nl}^{R}$ and turn to the
one-loop calculation of the renormalization constant $Z_{nl}$ and of the
critical dimension of the operator (\ref{Fnl}), which will be denoted as
$\Delta_{nl}$.

Let $\Gamma(x;\theta)$ be the generating functional of the
1-irreducible Green functions with one composite operator $F(x)=F_{nl}$
and any number of fields $\theta$. Here $x = \{ t,{\bf x}\}$ is
the argument of the operator and $\theta$ is
the functional argument, the ``classical analog'' of the random
field $\theta$. We are interested in the $n$-th term of the
expansion of $\Gamma(x;\theta)$ in $\theta$, which we denote
$\Gamma_{n}(x;\theta)$. It can be written as follows:
\begin{widetext}
\begin{equation}
\Gamma_{n}(x;\theta) =  \int dx_{1} \cdots \int dx_{n}
\, \theta(x_{1})\cdots\theta(x_{n})\,
\langle F(x) \theta(x_{1})\cdots\theta(x_{n})\rangle_{\rm 1-ir}.
\label{Gamma1}
\end{equation}
\end{widetext}
In the one-loop approximation the function (\ref{Gamma1}) is
represented diagramatically as follows:
\begin{equation}
\Gamma_{n}(x;\theta)= F(x) + \frac{1}{2}  \vcenter{\hbox
{\includegraphics [width=.09\textwidth,clip]{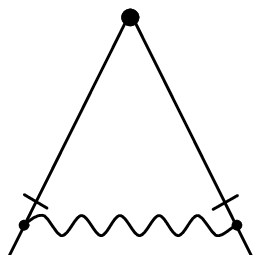}}}
\label{Gamma2}
\end{equation}
The first term is the tree (loopless) approximation, and the thick dot
with the two attached lines in the diagram denotes the operator vertex,
to be specified later.

The renormalization constant $Z_{nl}$ for the operator (\ref{Fnl})
is found from the requirement that the
renormalized analog $\Gamma_{n}^{R}=Z_{nl}^{-1}\Gamma_{n}$ of the function
(\ref{Gamma1}) be UV finite in terms of renormalized parameters.

For practical calculations, it is convenient to contract the tensors
(\ref{Fnl}) with an arbitrary constant vector
{\mbox{\boldmath $\lambda$}}$~=~\{\lambda_{i}\}$.
The resulting scalar operator has the form
\begin{equation}
F^{(n,l)} = (\lambda_{i}\theta _{i})^{l}
(\theta_{i}\theta_{i})^{s} + \dots,
\label{FnpSk}
\end{equation}
where the subtractions, denoted by the ellipsis, necessarily involve the
factors of $\lambda^{2}=\lambda_{i}\lambda_{i}$.

Within our accuracy, it is sufficient to replace all the renormalization
constants in the diagram with unities, so that $D_{0} \to g\nu^3\mu^y$,
$u_{0}\to u$ and so on. Furthermore, we are eventually interested in the
fixed-point value of the anomalous dimension, so that we can set $u=w=1$
in the following. Since the diagram is logarithmically divergent,
we can set all the external frequences and momenta equal to zero.

Like for the calculation of the self-energy diagram in sec.~\ref{sec:Sigma},
here and below we use numbers (instead of Latin letters) to denote the tensor
indices. Then the diagram in (\ref{Gamma2}) can be represented as follows:
\begin{equation}
V_{12}(\theta)\, C_{1278}\, \theta_{7}\theta_{8},
\label{VerTrans2}
\end{equation}
where $V_{12}(\theta)$ is the operator vertex (denoted by the thick dot
in the diagram and specified below),
the fields $\theta_{7}\theta_{8}$ (denoted by wavy tails) are attached to
the lower vertices (\ref{vertex1}) (small dots) and $C_{1278}$ is the
``core'' of the diagram. It has the form
\begin{equation}
C_{1278}=
\int \frac{d\omega}{2\pi}  \int_{k>m} \frac{d{\bf k}}{(2\pi)^{d}}
\frac  {g\nu^3\mu^y R_{1278}} { ({\omega^{2}+\nu^2 k^{4}})^{2}}
\label{Trian}
\end{equation}
with the tensorial factor stemming from the vertices (\ref{vertex1})
and the projectors of the propagators:
\begin{equation}
R_{1278}= P_{13}^{\bot}({\bf k}) P_{24}^{\bot}({\bf k})
\left\{ P_{56}^{\bot}({\bf k})+\alpha P_{56}^{\parallel}({\bf k})
\right\}
V_{375}({\bf k})V_{486}({\bf k}).
\label{TenStr}
\end{equation}

Intergation over the frequency is easily performed:
\begin{equation}
\int \frac{d\omega}{2\pi} \frac{1}{ ({\omega^{2}+\nu^2 k^{4}})^{2}}=
\frac{1}{4\nu^3k^{6}}.
\label{Frik}
\end{equation}

Contraction of the vector indices in (\ref{TenStr}) leads to the
following integrals over the momentum:
\begin{equation}
\int_{k>m} \frac{d{\bf k}}{(2\pi)^{d}} \frac{1}{k^{d+y}}\,
P_{12}^{\bot}({\bf k})\, P_{78}^{\parallel}({\bf k})
\label{Trod}
\end{equation}
for the transverse contribution in (\ref{lines}) and
\begin{equation}
\int_{k>m} \frac{d{\bf k}}{(2\pi)^{d}} \frac{1}{k^{d+y}}\,
P_{17}^{\bot}({\bf k})\, P_{28}^{\bot}({\bf k})
\label{Prod}
\end{equation}
for the longitudinal one. With the aid of the relations (\ref{tenz}), all
these integrals are reduced to the scalar integral (\ref{ska}).

Combining all these contributions and contracting the result with the
fields $\theta_{7}\theta_{8}$ gives for (\ref{VerTrans2}) the following
expression:
\begin{eqnarray}
\frac{-D_{0}}{4d(d+2)\nu^3}\, V_{12}\, \left\{ T_{12}+\alpha L_{12} \right\},
\label{Dyn}
\end{eqnarray}
where
\[ T_{12} = {(d+1)} \delta_{12}\theta^{2}  - 2 \theta_{1} \theta_{2} \]
and
\[ L_{12} = \delta_{12}\theta^{2} +(d^{2}-2) \theta_{1} \theta_{2}. \]

Now let us turn to the vertex factor
\begin{eqnarray}
V_{12} = \frac {\delta^{2} F_{nl}(x) } {\delta\theta_{1}(x_{1})
\delta\theta_{2}(x_{2})}.
\label{VerOp}
\end{eqnarray}
Using the chain rule, it can be rewritten in the form
\begin{eqnarray}
V_{12} = \frac{\partial^{2} F_{nl}(w) } {\partial w_{1}\partial w_{2}} \,
\delta(x-x_{1})\delta(x-x_{2}),
\label{VerOp3}
\end{eqnarray}
where
\begin{equation}
F^{(n,l)} = (\lambda_{i}w_{i})^{l} (w_{i}w_{i})^{s} + \dots,
\label{FnpSk2}
\end{equation}
with the subsequent substitution $w_{i} \to \theta_{i}(x)$.

The differentiation gives
\begin{widetext}
\begin{eqnarray}
{\partial^{2}F^{(n,l)}}/{\partial w_{1}\partial w_{2}} &=&
2s (w^{2})^{s-2} (\lambda w)^{l} \left[\delta_{12} w^{2} +2(s-1)w_{1}w_{2}
\right] + l(l-1) (w^{2})^{s}(\lambda w)^{l-2} \lambda_{1} \lambda_{2}+
\nonumber \\
&+& 2ls (w^{2})^{s-1}(\lambda w)^{l-1} (w_{1}\lambda_{2}+ w_{2}\lambda_{1}),
\label{Vae11}
\end{eqnarray}
\end{widetext}
where $w^{2}=w_{k}w_{k}$ and $(\lambda w)=\lambda_{k}w_{k}$.

Now we have to contract the vertex factor (\ref{Vae11}) with the expression
(\ref{Dyn}). In order to find the renormalization constant, it is sufficient
to retain only the terms proportional to the principal monomial in
(\ref{FnpSk}) and discard all the terms containing the factors of
$\lambda^{2}=\lambda_{i}\lambda_{i}$. Combining all the relevant factors
finally gives
\begin{eqnarray}
\Gamma_{nl} (x) =F_{nl} (x) \left\{ 1 - \left(\frac{\mu}{m}\right)^{y} \,
\frac{\hat g \left( Q_{1}+\alpha Q_{2} \right)}{8yd(d+2)}  \right\}\! ,
\label{Gamma_answ}
\end{eqnarray}
where
\begin{eqnarray}
Q_{1} &=& -n(n+d)(d-1) +(d+1) \, l(l+d+2) ,
\nonumber \\
Q_{2} &=& -n(n+nd-d)(d-1) + l(l+d+2)
\label{Q12}
\end{eqnarray}
and ${\hat g}$ is defined in  (\ref{ghat}).

The requirement that the renormalized analog of the function (\ref{Gamma1})
be UV finite in the MS scheme gives:
\begin{eqnarray}
Z_{nl}=\left\{1-\frac{\hat g}{8yd(d+2)}\, \left( Q_{1}+\alpha Q_{2} \right)
\right\},
\label{Znl}
\end{eqnarray}
and the anomalous dimension
$\gamma_{nl} = \widetilde{\cal D}_{\mu} \ln Z_{nl}$ is
\begin{eqnarray}
\gamma_{nl} = \frac{\hat g}{8d(d+2)}\, \left( Q_{1}+\alpha Q_{2} \right)
\label{Gnl}
\end{eqnarray}
(we recall that we already set $u=w=1$).

Using the general expression (\ref{Krit}), for the critical dimension
$\Delta_{nl}$ at the fixed point (\ref{FP}) we obtain
\begin{eqnarray}
\Delta_{nl} = n\Delta_{\theta}+\gamma_{nl}^{*} = \frac{ny}{6}+
\frac{y \left( Q_{1}+\alpha Q_{2} \right) } {6(d-1)(d+2)}
\label{Dnl}
\end{eqnarray}
with $\Delta_{\theta}$ from (\ref{KriTet}). In particular, for the scalar
operator one arrives at the expression
\begin{eqnarray}
\Delta_{n0} = \frac{-yn}{6(d+2)} \left\{ (n-2) +
\alpha \frac{(3n+d-4)}{(d-1)} \right\},
\label{Snl}
\end{eqnarray}
which is negative and decreases as $\alpha$ grows:
\begin{eqnarray}
\partial \Delta_{n0} / \partial \alpha <0.
\label{PRO}
\end{eqnarray}
As we will see in the next section, this means that the anomalous scaling
is indeed present in our model and becomes more strongly pronounced as
the degree of compressibility increases.

For a fixed $n$, the dimensions (\ref{Dnl}) exhibit a kind of hierarchy with
respect to the rank $l$ (which measures the ``degree of anisotropy''):
\begin{eqnarray}
\partial \Delta_{nl} / \partial l>0.
\label{hier}
\end{eqnarray}
In contrast to the Gaussian model (see, e.g., \cite{AKens,alpha}), this
hierarchy becomes more strongly pronounced as $\alpha$ increases:
\begin{eqnarray}
\partial^{2} \Delta_{nl} / \partial l\partial \alpha >0.
\label{hierA}
\end{eqnarray}

\section{Operator product expansion and the anomalous scaling}
\label{sec:OPE}

\subsection{General discussion and isotropic case}
\label{sec:OPEG}

The quantities of interest are, in particular, the pair correlation functions
of the (renormalized analogs of the) operators (\ref{Fnl}). In the following,
we restrict ourselves with the equal-time correlations because they are
Galilean invariant and do not bear strong dependence on the IR scale
$m=L^{-1}$ caused by the sweeping of small-scale vortices by the
large-scale ones. Then one can write
\begin{widetext}
\begin{eqnarray}
\langle F_{nl}(t, {\bf x}) F_{qj}(t, {\bf x}') \rangle =
 \mu^{d_{F}} \nu^{d^{\omega}_{F}} \eta_{nl,qj} (\mu r, mr, c/\mu\nu)
\simeq
\mu^{d_{F}} \nu^{d^{\omega}_{F}} (\mu r)^{-\Delta_{nl}-\Delta_{qj}}
\zeta_{nl,qj} (mr, c(r)),
\label{IRR}
\end{eqnarray}
\end{widetext}
where  $r=|{\bf r}|=|{\bf x}'-{\bf x}|$.
The first equality follows from simple dimensionality considerations; then
$d^{\omega}_{F}$ and $d_{F}$ are the canonical dimensions of the correlation
function, given by simple sums of the corresponding dimensions of the
operators, and $\eta(\dots)$ is some function of completely dimensionless
parameters. We have expressed the right-hand side in renormalized variables,
when the reference mass $\mu$ is the substitute of the UV momentum scale
$\Lambda$. The second (approximate) equality is valid in the IR asymptotic
range $\mu r \gg 1$ and follows from solving the RG equation in the presence
of the IR attractive fixed point; $\Delta_{nl}$ and $\Delta_{qj}$ being the
critical dimensions of the operators from the left-hand side, given by
Eq.~(\ref{Dnl}). In the following, we omit the RG invariant variable
$c(r) = c (\mu r)^{\Delta_{c}} / (\mu\nu)$, which is restricted in the IR
range; for more explanations, see~\cite{Kont}. We will also omit the indices
of the scaling functions $\eta$ and $\zeta$ and do not display the dependence
on the parameters $\nu$, $\mu$ that are fixed for a given physical setup.

The inertial range corresponds to the additional inequality $mr\ll1$. The
form of the functions $\zeta$ in (\ref{IRR}) is not determined by the RG
equations alone; they should be augmented by the OPE. In the case at hand,
the OPE states that the equal-time product $F_{nl}(x)F_{qj}(x')$
at ${\bf x} = ({\bf x} + {\bf x'} )/2 = {\rm const}$ and
${\bf r} = {\bf x}' - {\bf x}\to 0$ can be represented in the form
\begin{equation}
F_{nl}(x)F_{qj}(x') \simeq \sum_{F} C_{F} ({\bf r}) F(t,{\bf x}).
\label{OPE}
\end{equation}
Here the functions $C_{F}$ are the Wilson coefficients, regular in $m^{2}$,
and $F$ are, in general, all possible renormalized local composite operators
allowed by the symmetry of the model and of the left-hand side. In the case
at hand this implies that only Galilean invariant operators contribute.
If these operators have additional vector indices, they are contracted with
the corresponding (additional) indices of the coefficient functions $C_{F}$.

Without loss of generality, it can always be assumed that the expansion
(\ref{OPE}) is made in the irreducible operators with definite critical
dimensions $\Delta_{F}$. The correlation functions (\ref{IRR}) are obtained
by averaging the expression (\ref{OPE}) with the weight
$\exp {\cal S}(\Phi)$, where ${\cal S}(\Phi)$ is the (renormalized) action
functional (\ref{FactR}). Then the quantities $\langle F \rangle$ appear on
the right-hand sides. Consider first the isotropic case, then only the
contributions from scalar operators survive.
Their asymptotic behavior for $m\to0$ is found from the RG equations for the
operators $F$ and has the form $\langle F \rangle \propto  m^{\Delta_{F}}$
(we recall that $\Delta_{m}=1$).

Thus, combining the expressions (\ref{IRR}) and (\ref{OPE}) gives the
following inertial-range asymptotic representation for the scaling
functions $\zeta$:
\begin{equation}
\zeta(mr) \simeq \sum_{F} A_{F}\, (mr)^{\Delta_{F}},
\label{OR}
\end{equation}
where all the coefficients $A_{F}=A_{F}(mr)$ are regular in $(mr)^{2}$.

Singularities for $mr\to0$ (and thus the anomalous scaling) result from the
contributions in (\ref{OR}) of the operators with {\it negative} critical
dimensions, termed ``dangerous'' in \cite{turbo}. Clearly, if the number of
such operators were finite, the leading contribution would be determined
by the operator with the lowest dimension. However, one can argue that, if
at least one dangerous operator exists in a model, the latter necessarily
involves an infinite set of dangerous operators, and the spectrum of their
dimensions is not bounded from below; see Appendix~A for discussion.
In our case, from the expression (\ref{Snl}) we can see that all
the scalar operators $F_{n0}$ are dangerous, and their dimensions
$\Delta_{n0}$ increase without bound as $n$ grows.

Fortunately, the linearity of the original equation (\ref{mhd}) in the field
$\theta$ imposes the restriction that the number of the fields $\theta$ in
all the composite operators in the expansion (\ref{OPE}) cannot exceed their
number in the left-hand side; cf. the remark below Eq.~(\ref{indexo}) in
Sec.~\ref{sec:Opera}. In turn, this means that only a finite number of the
operators of the type $F_{k0}$ can contribute to any given OPE. For the
product (\ref{OPE}), these are the operators with $k\le n+q$. Thus,
\begin{equation}
\zeta(mr) \simeq \sum_{k=0}^{n+q} A_{k} (mr)\, (mr)^{\Delta_{k0}} + \dots
\label{Fin2}
\end{equation}
with $\Delta_{k0}$ from (\ref{Snl}); the ellipsis stands for the ``more
distant'' corrections to the small-$mr$ behavior, given by the operators
with derivatives and other types of fields. The leading term in (\ref{Fin2})
is determined by the operator with the maximum possible $k=n+q$, so that
the final leading-order asymptotic expression for the correlation function
(\ref{IRR}) in the inertial range $\mu r\gg1$, $mr \ll 1$ has the form
\begin{equation}
\langle F_{nl} F_{qj}\rangle \simeq
(\mu r)^{-\Delta_{n}-\Delta_{q}} (mr)^{\Delta_{n+q}}.
\label{FinF}
\end{equation}
As already mentioned in the preceding section, the inequality (\ref{PRO})
means that the anomalous scaling becomes more strongly pronounced as the
degree of compressibility grows. We also note that the inequality
\[ \Delta_{n+q} < \Delta_{n}+\Delta_{q}, \]
which follows from the explicit expressions (\ref{Dnl}) and, in fact,
is required
by the probabilistic theory, shows that the expression (\ref{FinF})
diverges for $r\to0$.

\subsection{Effects of large-scale anisotropy}
\label{sec:Antz}

Consider effects of the anisotropy, introduced into the system at large
scales $\sim L$ through, say, the large-scale field $B^{0}_{i}=n_{i} B^{0}$
or through the correlation function of the artificial random force.
Then the irreducible tensor composite operators acquire nonzero mean values,
built of the vector ${\bf n}$: for example, the mean value of the second-rank
operator is proportional to the irreducible tensor $n_{i}n_{j}-\delta_{ij}/d$.
In general, the mean value of any $l$-th rank irreducible operator is
proportional to the tensor $n_{i_{1}}\dots n_{i_{l}} + \dots$, where the
ellipsis stands for the appropriate subtractions with the Kronecker $\delta$
symbols that make it irreducible. Upon substitution into the OPE for the
product of two scalar operators, their tensor indices are contracted with the
corresponding indices of the coefficient functions $C_{F}({\bf r})$. This
gives rise to the Gegenbauer polynomials, the $d$-dimensional analogs of the
Legendre polynomials $P_{l}(\cos\vartheta)$, where $\vartheta$ is the angle
between the vectors ${\bf r}$ and ${\bf n}$. For general anisotropy,
the $d$-dimensional spherical harmonics appear on the right-hand side,
while for products of tensor operators, their tensor analogs arise.

Consider, as the simplest example, the pair correlation function (\ref{IRR})
of two scalar operators in the inertial range:
\begin{eqnarray}
\langle F_{n0}(t, {\bf x}) F_{q0}(t, {\bf x}') \rangle \simeq
r^{-\Delta_{n0}-\Delta_{q0}} \times
\nonumber \\ \times
\sum_{l=0}^{N} c_{l} P_{l}(\cos \vartheta)
(mr)^{\Delta_{Nl}} + \dots
\label{shell}
\end{eqnarray}
with $N=n+q$ and $\Delta_{Nl}$ from (\ref{Dnl});
$c_{l}$ are numerical coefficients and the ellipsis stands for the
``distant'' contributions with $l>N$. The inequality (\ref{hier}) means that
the anisotropic contributions in (\ref{shell}) exhibit a kind of hierarchy
related to the degree of anisotropy $l$: the leading contribution is given by
the isotropic ``shell'' ($l=0$), while the contributions with $l>1$ give only
corrections which become relatively weaker as $mr\to0$, the faster the higher
the degree of anisotropy $l$ is. Similar hierarchy, observed earlier in
numerous models, e.g.,
\cite{Lanotte,ABP,Lanotte2,A99,AKens,AK2,alpha,amodel,TWP,Kont},
gives quantitative support for Kolmogorov's hypothesis of the local isotropy
restoration.

The inequality (\ref{hierA}) means that the hierarchy (\ref{hier}) becomes
stronger as the degree of compressibility $\alpha$ grows: the anisotropic
corrections are getting further from one another and from the isotropic term,
contrary to the situation observed earlier for passive vector field, advected
by Kraichnan's ensemble \cite{alpha}. A similar discrepancy for the scalar
field was encountered recently in~\cite{Kont}. This means that the results
obtained on the base of simplified ``synthetic'' ensembles should be taken
with some precaution.

\section{Conclusion} \label{sec:Conc}

We have studied the model of passive vector field, advected by a turbulent
flow. The latter is described by the Navier--Stokes equations for a strongly
compressible fluid (\ref{ANU}), (\ref{ANU1}) with an external stirring force
with the correlation function $\propto k^{4-d-y}$; (\ref{force}),
(\ref{power}). From physics viewpoints, the model describes
magnetohydrodynamic turbulence in the so-called kinematic approximation,
where the effects of the magnetic field on the dynamics of the fluid are
neglected.

The full stochastic problem can be cast as a field theoretic model
with the action functional specified in (\ref{FSA}) and (\ref{MHDA}).
That model appears multiplicatively renormalizable, so that
the corresponding RG equations can be derived in a usual way.
They have the only IR attractive fixed point in the physical range of
parameters, so that various correlation functions reveal scaling behavior
in the IR region.

Their inertial-range behavior was studied by means of the OPE; existence
of anomalous scaling (singular power-like dependence on the integral scale
$L$) was established. The corresponding anomalous exponents were identified
with the scaling (critical) dimensions of certain composite fields
(composite operators), namely, powers of the magnetic field. They can be
systematically calculated as series in the exponent $y$. The practical
calculation was accomplished in the leading order; the results are presented
in (\ref{Dnl}).

The results obtained are quite similar to those derived earlier for the
vector fields advected by synthetic velocity ensembles \cite{alpha,TWP}:
the anomalous scaling becomes more remarkable as the degree of
compressibility $\alpha$ increases; the anisotropic contributions form an
hierarchy related to the degree of anisotropy $l$, so that the leading
inertial-range contribution is the same as for the isotropic case.
However, that hierarchy becomes stronger as the degree of compressibility
grows, in contrast to what was observed in~\cite{alpha} for the Kraichnan's
rapid-change ensemble. In this respect, our results are close to
what was recently observed for the scalar field, advected by the same
velocity ensemble \cite{Kont}.

From a more theoretical point of view, it is important that in our case, the
anomalous exponents are associated with the critical dimensions of certain
individual composite operators, exactly as in the RG+OPE treatment of the
rapid-change models; see, e.g., \cite{JphysA,Lanotte2,alpha,RG,cube}.
In the zero-mode approach to the latter, the anomalous exponents are related
to the so-called zero modes (unforced solutions) of the exact differential
equations satisfied by the equal-time correlation functions; see, e.g.,
\cite{V96,RK97,Lanotte,FGV}. In a more general sense, zero modes can be
interpreted as certain statistical conservation laws in the dynamics of
particle clusters \cite{FGV}.
The close resemblance in the RG+OPE pictures of the origin of anomalous
scaling for the present model and its rapid-change predecessors suggests
that the concept of zero modes (and thus that of statistical
conservation laws) is also applicable in much more realistic models.

\appendix
\section{Infinite number of negative dimensions}

Let $F({\bf x})$ be a certain renormalized composite operator in a certain
multiplicatively renormalizable field theoretic model with an IR attractive
fixed point of the RG equations. Assume that $F$ has a definite negative
critical dimension, $\Delta_{F}<0$, and assume that it is the lowest
dimension in the model (that is, $F$ is the ``most dangerous'' operator).

Consider its pair correlation function:
\begin{equation}
\langle F({\bf x})F({\bf x}') \rangle = \int \frac{d{\bf k}}{(2\pi)^{d}}\,
D_{F}(k) \exp \{ {\rm i} {\bf k} \cdot ({\bf x}'-{\bf x}) \},
\label{Korrr}
\end{equation}
where $k=|{\bf k}|$. Our assumption implies that in the IR range the
function $D_{F}(k)$ has the asymptotic form
\begin{equation}
D_{F}(k) \simeq k^{-d+2+ \Delta_{F}} f (m/k),
\label{Ass}
\end{equation}
where $m$ is some typical IR momentum scale, $f(m/k)$ is some dimensionless
scaling function, and we assumed that $\Delta_{m}=1$ (like in our model).
Now consider the mean value
\begin{equation}
\langle F^{2}({\bf x}) \rangle = \int \frac{d{\bf k}}{(2\pi)^{d}}\,
D_{F}(k).
\label{Korr2}
\end{equation}
The function $f(m/k)$ provides IR regularization of the integral
(\ref{Korr2}). The question is whether this integral remains convergent
for large $k$ if the exact (unknown) function $D_{F}(k)$ is replaced with
its asymptotic form (\ref{Ass}). According to the OPE, the asymptotic
behavior of the function $f(m/k)$ for large $k$, or, equivalently, for
small $m$ is determined by the contribution of the most dangerous
operator, which by assumption  is $F$ itself:
\begin{equation}
f (m/k) \simeq (m/k)^{\Delta_{F}}.
\label{Ass2}
\end{equation}
Thus, for large $k$ we have
\begin{equation}
D_{F}(k) \simeq k^{-d+\Delta_{F}}
\label{Assj}
\end{equation}
and the integral (\ref{Korr2}) remains convergent upon the substitution of
(\ref{Ass}). In turn, this means that it is expressed only in terms of the
IR parameter $m$ (UV regularization $\Lambda$ can be removed). Then it is
easily found from the dimension:
\begin{equation}
\langle F^{2}({\bf x}) \rangle \simeq m^{2\Delta_{F}}.
\label{Korr3}
\end{equation}

Expression (\ref{Korr3}) means, however, that the operator $F^{2}$ has the
negative dimension $2\Delta_{F}<\Delta_{F}<0$ which is smaller than that of
$F$. We arrive at the contradiction with our initial assumption about $F$.

To avoid possible misunderstanding we stress that our consideration does not
mean that $F^{2}$ is necessarily dangerous and its dimension is exactly
$2\Delta_{F}$ (although this indeed happens, e.g., for the powers of the
velocity field in the stochastic NS problem; see \cite{turbo}). It means
that operators with negative dimensions, if any, always appear in a model
as infinite families, with the spectrum of dimensions not bounded from below.
This fact should be taken into account in axiomatic or phenomenological
implementations of the OPE to models of turbulence~\cite{FZ}.

\section*{Acknowledgments}

The authors are indebted to L.~Ts.~Adzhemyan, Michal~Hnatich, Juha~Honkonen,
and M.~Yu.~Nalimov for discussion.

The Authors acknowledge the Saint Petersburg State University for the
research grant 11.38.185.2014.

Mariia Kostenko was also supported by the Dmitry Zimin's Dynasty Foundation.


\end{document}